\documentclass[twocolumn,prb,longbibliography,superscriptaddress]{revtex4}
\usepackage{amsmath,amssymb}
\usepackage[english]{babel}
\usepackage[T1]{fontenc}
\usepackage{inputenc}
\usepackage{graphicx}
\usepackage{xcolor}
\usepackage{float}
\usepackage{multirow}
\usepackage{gensymb}
\usepackage[colorlinks]{hyperref}
\hypersetup{
  linkcolor  = blue,
  citecolor  = blue,
  urlcolor   = blue,
  colorlinks = true,
}

\begin{document}

\title{\textbf{Metastability of the Topological Magnetic Orders in the Chiral Antiferromagnet EuPtSi}}

\author{S.~Rousseau}
\affiliation{Univ. Grenoble Alpes, CEA, Grenoble INP, IRIG, PHELIQS, F-38000 Grenoble, France}
\affiliation{Laboratoire National des Champs Magn\'etiques Intenses (LNCMI), CNRS, Univ. Grenoble Alpes, 38042 Grenoble, France}
\author{G.~Seyfarth}
\affiliation{Laboratoire National des Champs Magn\'etiques Intenses (LNCMI), CNRS, Univ. Grenoble Alpes, 38042 Grenoble, France}
\author{G.~Knebel}
\affiliation{Univ. Grenoble Alpes, CEA, Grenoble INP, IRIG, PHELIQS, F-38000 Grenoble, France}
\author{D. Aoki}
\affiliation{Institute for Materials Research, Tohoku University, Ibaraki 311-1313, Japan}
\author{Y.  \=Onuki}
\affiliation{RIKEN Center for Emergent Matter Science, Wako, Saitama 351-0198, Japan}
\author{A.~Pourret}
\email[E-mail me at: ]{alexandre.pourret@cea.fr}
\affiliation{Univ. Grenoble Alpes, CEA, Grenoble INP, IRIG, PHELIQS, F-38000 Grenoble, France}

\date{\today} 

\begin{abstract}

We report resistivity and Hall effect measurements in the chiral antiferromagnet EuPtSi. Depending on the magnetic field orientation with respect to the crystallographic axes, EuPtSi presents different topological magnetic phases below the Néel temperature $T_N=4.05$~K. In particular, for a field $H \parallel $ [111], it exhibits  the well known skyrmion lattice A-phase inside the conical phase between $T=0.45$~K and $T_N$ in the field range from 0.8~T to 1.4~T. Remarkably, the skyrmion lattice state in EuPtSi, composed of nanoscale skyrmions, can be extended down to very low temperature (lower than 0.1~K) through field-cooling regardless of the cooling rate and of the magnetic history. Similarly the metastability of the A'- and B-phases ($H \parallel $ [100]) at low temperature is evidenced by our measurements. These results suggest that EuPtSi is a peculiar example where the competition  between the topological stability and the thermal agitation can lead to metastable quantum skyrmion state.
\end{abstract}


\maketitle

\section{Introduction} \label{sec:intro}

In the past decade, the role of topological properties in condensed matter physics has been uncovered in many different systems: 2D electron gas in the quantum Hall regime, topological insulators, and Weyl or Dirac semi-metals, topological superconductors.\cite{Chang_RevModPhys_2023, Hasan_Kane_RevModPhys_2010, Sato_2017} Topological spin textures have also been investigated, stimulated by the possibility of discovering unusual physical phenomena owing to the interplay between magnetism and topology.\cite{Zhou_2025}  A prominent example is the magnetic skyrmion, a non-collinear spin structure with particle-like topologically protected states, which give it enormous stability even at small sizes and which make it a potential carrier of information in future data storage devices, such as racetrack nano devices.\cite{fert_advances_2017,zhang_chiral_2018,everschor-sitte_perspective_2018} In most systems experimentally investigated to date, skyrmions emerge as classical objects forming a skyrmion lattice (SkL).\cite{tokura_magnetic_2021} However, the discovery at very low temperature of skyrmions with nanometer length scales (spins wound over a few lattice spacings only), where quantum effects cannot be ignored has sparked interest in their quantum properties, introducing the notion of quantum skyrmions, such as quantum tunneling and energy-level quantization.\cite{Lohani_PRX2019, Siegl_PhysRevResearch2022, Salvati_PRB2024}

Experimental observation of magnetic SkL has been primarily limited to a narrow, finite temperature region in which the skyrmion phase is thermodynamically stable. Recently, metastable SkL with long-lived skyrmions have been realized in many materials, extending by far the thermodynamically stable skyrmion phase. Various processes have been used to obtain this metastable SkL state: rapid temperature quenching in the prototypical material MnSi
,\cite{nakajima_skyrmion_2017} electrical field in Cu$_2$OSeO$_3$
,\cite{okamura_transition_2016} specific magnetic field cooling in  Co$_8$Zn$_8$Mn$_4$
,\cite{Karube2016, morikawa_deformation_2017} or laser heating in cobalt-based trilayers.\cite{olleros-rodriguez_non-equilibrium_2022}

Recently, metastability has been also observed in the SkL A-phase of the chiral anti-ferromagnet EuPtSi by field-cooling the system.\cite{sakakibara_magnetic_2021} EuPtSi crystallizes in a non-centrosymmetric cubic chiral structure with space group $P2_1 3$ ($T^4$),\cite{kakihana_split_2015, onuki_unique_2020,kakihana_unique_2017} similar to MnSi and other B20 materials. The divalent Eu ions form a three-dimensional lattice of corner-sharing triangles called the trillium lattice.\cite{franco_fluctuation-induced_2017} This particular crystal structure gives rise to a magnetically highly frustrated ground state.\cite{Hopkinson2006} EuPtSi presents different topological magnetic orders depending on the magnetic field orientation below the ordering temperature $T_N=4.05$~K.\cite{takeuchi_angle_2020} For magnetic field $H \parallel [111]$, it exhibits the SkL A-phase inside the conical phase between
$T = 0.45$~K and $T_N$ in the field range from 0.8 T to 1.4 T. \cite{kakihana_giant_2018} This has been demonstrated by the observation of a triple-$\mathbf{q}$ magnetic ordering.\cite{kaneko_unique_2019,tabata_magnetic_2019, Matsumura_PRB_2024}
For $H \parallel$~[100], two topological magnetic orders appear in the conical phase, the A'-phase and the B-phase. The microscopic determination of the magnetic ordering vectors of the A' and B-phases is still missing (double-$\mathbf{q}$, triple-$\mathbf{q}$ or multi-$\mathbf{q}$). \cite{sakakibara_magnetic_2021,takeuchi_magnetic_2019} For $H \parallel$~[110], no exotic magnetic orders have been reported inside the conical phase. The strong anisotropy of the magnetic structures in EuPtSi has been linked to the small length $\lambda=18$~\r{A} of the magnetic ordering vector, ten times smaller than in MnSi, indicating the strong coupling of the helical state to the underlying crystal lattice.\cite{hayami_field-direction_2021} The short period explains the strong antiferromagnetic behavior of EuPtSi compared to the nearly ferromagnetic behavior of MnSi in which neighboring moments are nearly collinear.\cite{muhlbauer_skyrmion_2009,kakihana_unique_2019}

In this paper, we present a detailed study of the different field induced magnetic orders in EuPtSi by resistivity and Hall effect measurements at low temperature. We confirm the field cooled metastability for the skyrmion A-phase for $H \parallel $ [111] and observe also metastability for the exotic A' and B-phases for $H \parallel $ [100]. This evidences strong similarities between these three phases even if the presence of a SkL in the A' and B-phases has not been microscopically confirmed yet. 

\section{Experimental details}

The single crystals of EuPtSi were grown by the Bridgman method, details on the crystal growth can be found in Ref.~\onlinecite{kakihana_unique_2017}. All samples are thin slabs with dimensions of approximately $1.6~$mm$\, \times\, 0.5$~mm$\,\times\, 0.1$~mm  with the largest flat face normal to the applied magnetic field and were oriented using Laue diffractometer. The resistivity measurements were performed with the common AC 4-point method at low temperatures down to 100~mK and applied magnetic fields up to $H_{app}=16~$T. For the Hall effect, measurements under positive and negative magnetic fields have been carried out to correct the misalignment of the contacts. The maximum applied current  has been 0.5~mA. 
The angular dependence of the magnetoresistance has been measured at 1.75~K in a Quantum Design Physical Property Measurement System with a maximum field of 9~T. 

In all the shown data, $H$ refers to the effective magnetic field that differs from the applied magnetic field $H_{app}$ due to a large demagnetizing field in EuPtSi (the magnetization $M$ reaches 7~$\mu_B /$Eu at 3~T) and $H = H_{app} - NM$. The demagnetization factor has been estimated by \[N \approx \frac{4ab}{4ab + 3c(a+b)} \] for a sample with dimensions $2a\times 2b\times 2c$ and the field $H \parallel c$.\cite{prozorov_effective_2018}

\section{Results and discussion } \label{sec:results}

\subsection{Magnetic field applied along [111]}

\begin{figure}[h]
    \begin{center}
    \includegraphics[width=0.48\textwidth]{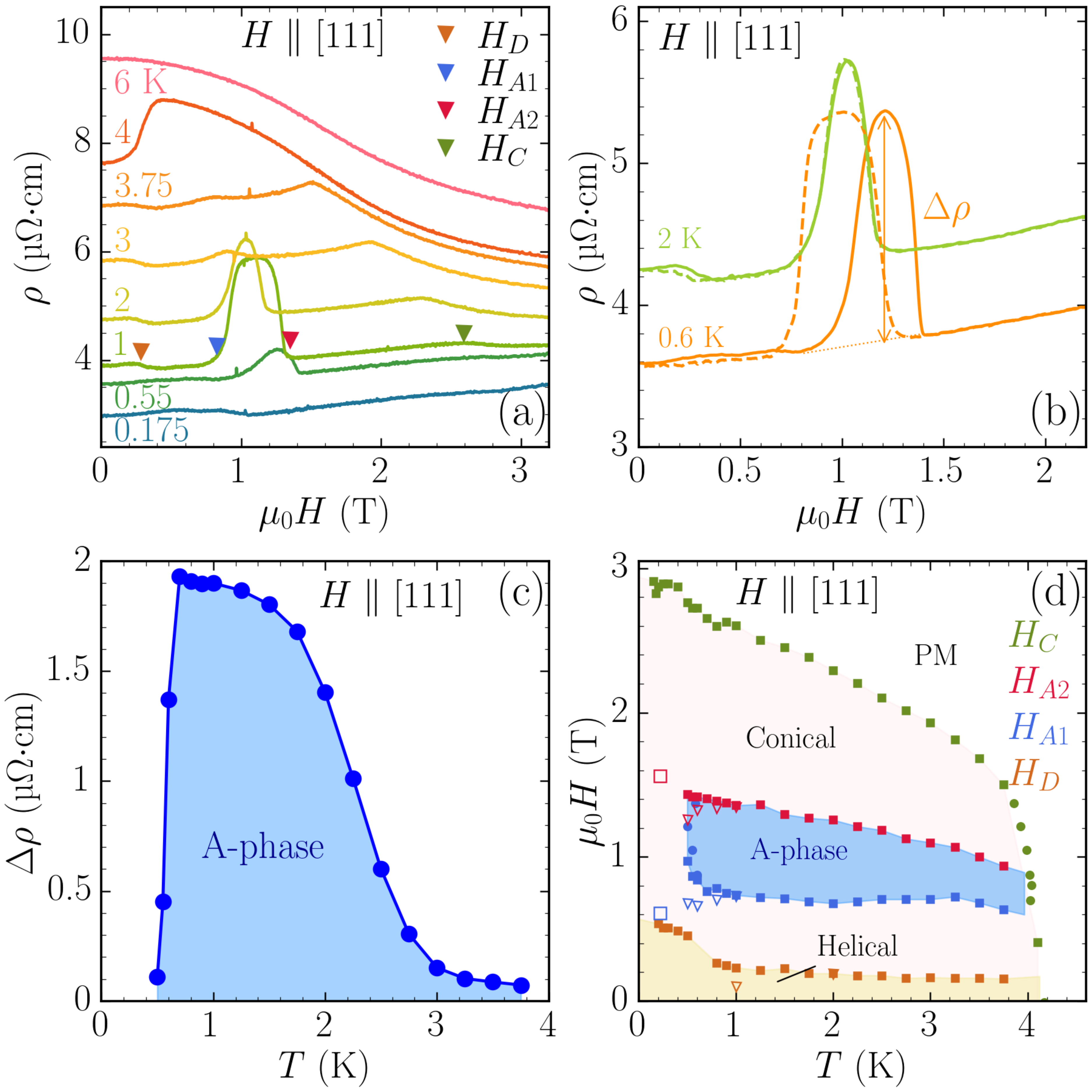}
    \caption{(a) Field dependence of $\rho$ at various temperatures above 0.175~K, for increasing magnetic field sweeps between 0~T and 3~T along the [111] direction with the current $J \parallel  [\Bar{1}\Bar{1}2]$. The 0.175~K curve is shifted vertically by $-0.5~\mu\Omega\cdot$cm for clarity. The magnetic transitions are indicated by colored triangles on the 1~K data. (b) Comparison between increasing (full lines) and decreasing (dashed lines) $\rho(H)$ field sweeps for $T = 2$ and 0.6~K. The arrow indicates the jump $\Delta\rho$ at 0.6~K. (c) The amplitude of the jump $\Delta\rho$ in the resistivity as a function of the temperature. (d) Magnetic phase diagram obtained for the [111] direction by resistivity measurements. Full squares (open triangles) are transition fields from increasing (decreasing) field sweeps. Circles denote transitions obtained by increasing temperature sweeps of $\rho$. The two open squares correspond to an extrapolation at 0.22~K of the phase boundaries / respective critical field lines of the thermodynamic equilibrium A-phase ($H_{A1}$ and $H_{A2}$), see text about the metastable SkL regime and also Fig.~\ref{Fig2}~(b).}
    \label{Fig1}
    \end{center}
\end{figure}

Fig. \ref{Fig1}~(a) shows the field dependence of the resistivity $\rho (H)$ of EuPtSi for field $H$ applied along the [111] direction and current $J \parallel [\Bar{1}\Bar{1}2] \perp H$ for a wide range of temperatures from 0.175~K to 6~K and for magnetic fields between 0 and 3~T. In this field range, at 6~K, the resistivity decreases smoothly with magnetic field. At 4~K, just below the magnetic ordering temperature, the resistivity increases strongly around 0.3~T and then decreases towards higher fields. As will be shown below in the phase diagram [see Fig.~\ref{Fig1}~(d)] this initial increase corresponds to the transition from the magnetically ordered state to the paramagnetic state above a critical field $H_C$. At lower temperatures, four transitions, indicated by colored triangles on the curve for $T = 1$~K, can be observed and tracked down to 0.5~K. At 1~K, the first transition $H_D$ occurs around 0.25~T as a downward step and corresponds to the transition from the helical ordered state to the conical state. The zero-field cooled (ZFC) A-phase appears as a large bump between the transitions $H_{A1}$ at 0.8~T and $H_{A2}$ at 1.3~T. Fig.~\ref{Fig1}~(b) shows the field dependence $\rho(H)$ at 2~K and 0.6~K.  The amplitude of $\Delta\rho (H)$ in the A-phase is marked for 0.6~K by the vertical arrow. Fig.~\ref{Fig1}~(c) shows the temperature dependence of $\Delta\rho$, the additional scattering contribution related to the A-phase. It starts to increase significantly below 3~K and is maximum at 0.65~K. To lower temperatures it decreases rapidly and $\Delta\rho \sim 0$ below 0.45~K. The amplitude of the resistivity jump $\Delta\rho$ when entering the A-phase, see Fig.~\ref{Fig1}~(c), allows us to clearly delimit the zero-field cooled (ZFC) A-phase. 
Above 3.75~K, the A-phase transitions become indistinguishable from $H_C$, the transition from the conical state to the field-polarized state, as $H_C$ rapidly drops to 0~T close to $T_N = 4.05$~K. 
Our resistivity measurements confirm the first-order nature of the $H_{A1}$ and $H_{A2}$ transitions below $T=1$~K, as shown in Fig.~\ref{Fig1}~(b), where a strong hysteresis of $\Delta \rho$ between increasing and decreasing field curves occurs. As the temperature decreases below 1~K, the transitions corresponding to the A-phase reach lower fields for decreasing field sweeps. Above 1~K, no further hysteretic feature was observed in the resistivity, although other studies have reported the first-order nature of all transitions \cite{franco_fluctuation-induced_2017,sakakibara_fluctuation-induced_2019,sakakibara_magnetic_2021,tabata_magnetic_2019}.

 Fig.~\ref{Fig1}~(d) depicts the magnetic phase diagram of EuPtSi for field applied along the [111] direction, determined from resistivity measurements. 
 The colors of the transition fields and temperatures, respectively indicated by full squares and circles, correspond to the four observed transitions, namely $H_D$, $H_{A1}$, $H_{A2}$ and $H_C$. Open triangles denote the magnetic hysteretic features of the A-phase transitions.
 
\begin{figure}[th!]
\begin{centering}
\includegraphics[width=0.48\textwidth]{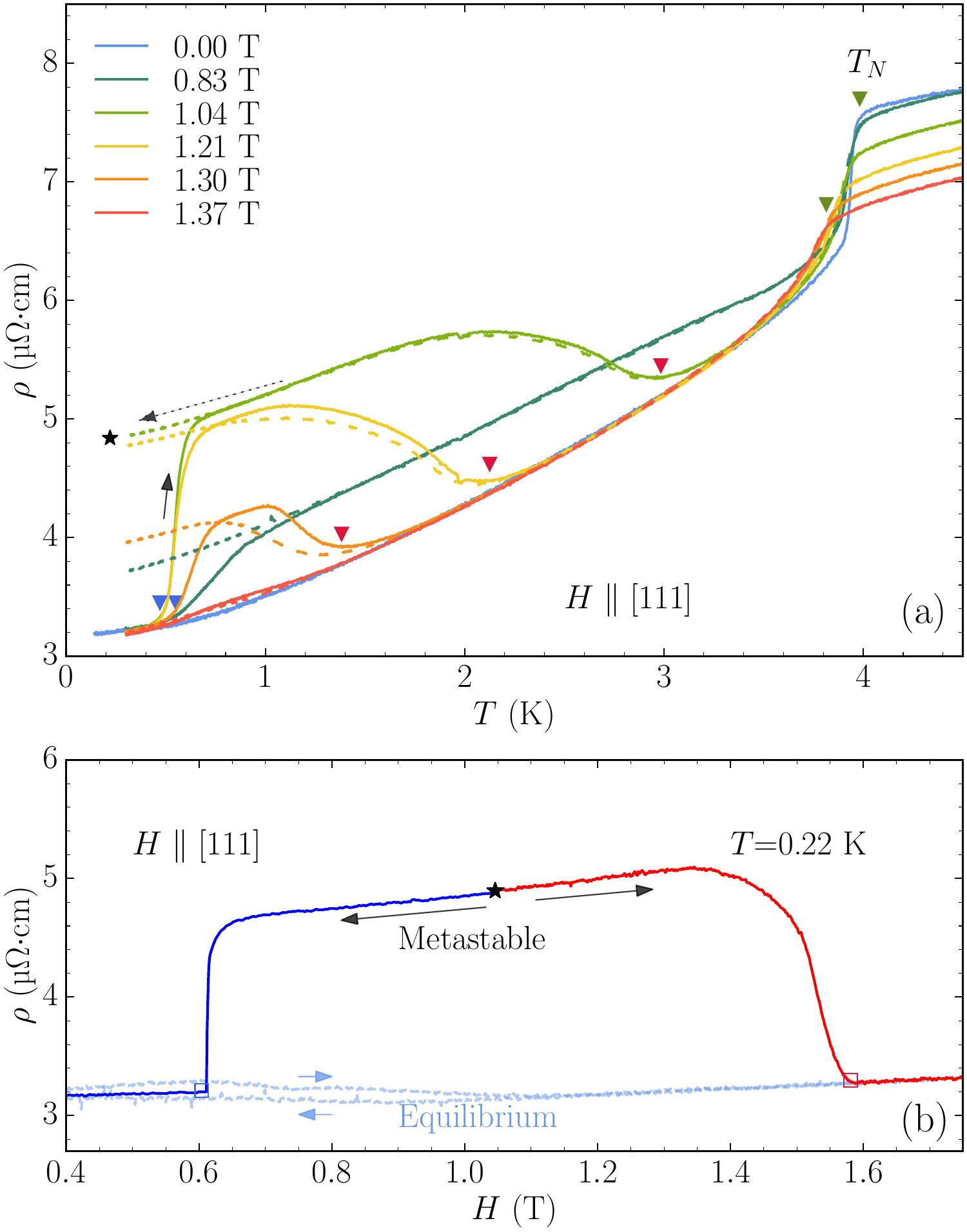}
\caption{(a) Temperature dependence $\rho(T)$ for various fixed magnetic fields along the [111] and current $J \parallel [\Bar{1}\Bar{1}2]$ after initial zero field cooling (ZFC). Solid lines (dashed lines) denote increasing (ZFC) (decreasing, (field cooled (FC)) temperature sweeps between 0.25 K and 4.5 K. Arrows indicate the variation of the temperature for the 1.37~T curve. The blue triangles indicate the entering into the A-phase on heating, and red triangles indicate transitions from the A-phase to the conical state. Green triangles denote $T_N$ at 0~T and 1.37~T). The black star indicates the metastable starting point in panel (b).  
(b) Field sweeps performed from the metastable state at $T= 0.22$~K, with the starting point at $H =1.04$~T indicated by a (black) star. Open blue (red) squares indicate the sharp transitions back to the conical state on decreasing (increasing) the field from the starting point. The dashed line shows the field sweep after zero field cooled at $T = 0.175$~K for comparison (blue arrows indicate the direction of the field).}
\label{Fig2}
\end{centering}
\end{figure}

 The temperature dependence of the resistivity $\rho(T)$ is presented in Fig.~\ref{Fig2}~(a) for various fixed magnetic fields applied along the [111] direction and with the current $J \parallel [\Bar{1}\Bar{1}2]$. For each field measurement, the magnetic field is applied at low temperature after zero field cooling (ZFC), keeping the system in the helical/conical state. The data were then acquired by sweeping the temperature continuously from 0.25~K to 6~K  field heated (FH), and then back to low temperature field cooled (FC). At zero field $\rho(T)$ shows only a step-like decrease when entering the helical phase at $T_N$ = 4.05~K indicating the ordering of the spins and therefore a reduction of magnetic scattering of the conduction electrons. This transition, corresponding to $H_C$ for field sweeps, decreases to lower temperatures with increasing field and fits very well the critical line $H_C (T)$ in the previous phase diagram (Fig.~\ref{Fig1}~(d)). 

 For magnetic fields above the lower limit of the A-phase shown in Fig.~\ref{Fig1}(d), with increasing temperature a transition to the A-phase occurs.  
 At $H = 0.83$~T, the transition is broadened, ranging from 0.45~K to 1.1~K. At this field the A-phase is stable up to the magnetic transition at $T_N$. At higher fields, see curves at 1.04~T and 1.21~T, the transition to the A-phase is rather sharp, and at higher temperatures a transition back to the conical state occurs below the transition to the paramagnetic state. At the transition from the A-phase to the conical phase [see phase diagram in Fig.~\ref{Fig1}(d)] a hysteresis between the heating and cooling curves occurs. This hysteresis is most pronounced for the curve at 1.3~T and signals the first order nature of the transition. 

 The most important observation is the strong hysteresis between the zero-field cooled  field-heated (ZFC-FH) and field-cooled (FC) curves at low temperatures. The ZFC-FH curve shows the entrance in the SkL A-phase, with a significant increase of the resistivity of up to 1.8~$\mu\Omega\cdot$cm, while when field cooling, no transition back to the conical phase is observed and the SkL A-phase is stable down to the lowest temperatures as a metastable  phase [see dashed lines in Fig.~\ref{Fig2}(a)]. 
 In ZFC conditions, the SkL phase is stabilized by thermal fluctuations at finite temperatures above 0.45~K which is very low compared to $T_N = 4.05$~K (around 10\%). In FC conditions the SkL phase survives to even lower temperatures, down to the lowest accessible ones by our experiment, so that we can not conclude on the existence nor the magnitude of a putative lower limit in that case. Anyhow, in the free energy landscape of the different spin configurations of EuPtSi the local minimum of the A-phase is only slightly different from that of the conical state.  
 
 
Next we investigate the metastable regime at low temperatures. After attaining the metastable SkL state by field cooling at a field of $H=1.04$~T down to $T$= 0.22~K, up and down field sweeps $\rho(H\nearrow,H\searrow)$ are shown in Fig.~\ref{Fig2}~(b). Starting from the same point in this metastable state ($T = 0.22$~K and $H = 1.04$~T), both field sweeps exhibit a sharp transition to the conical state. The critical fields delimiting the metastable state (open squares, at about 0.6 and 1.6~T) correspond to the extrapolation at 0.22~K of the respective critical field lines of the thermodynamic equilibrium A-phase ($H_{A1}$ and $H_{A2}$), as depicted by open squares on the phase diagram in Fig.~\ref{Fig1}~(d). The metastable A-phase is strongly irreversible against a field sweep. Subsequent field sweeps can not recover the metastable state, but remain in the equilibrium state. The complete field-cooling procedure has to be repeated in this case.
The metastable A-phase state can be obtained by field-cooling indifferently of the magnetic history and whether the system is first heated at zero field or under field. We verified that the metastable state is independent on the cooling rate. We have also confirmed the stability of this metastable state with respect to time, with no observable change in $\rho$ for at least up to 72~hours, emphasizing the extremely slow dynamic of the supercooled A-phase.

\subsection{Magnetic field applied along [100]}

\begin{figure}[th!]
\begin{center}
\includegraphics[width=0.48\textwidth]{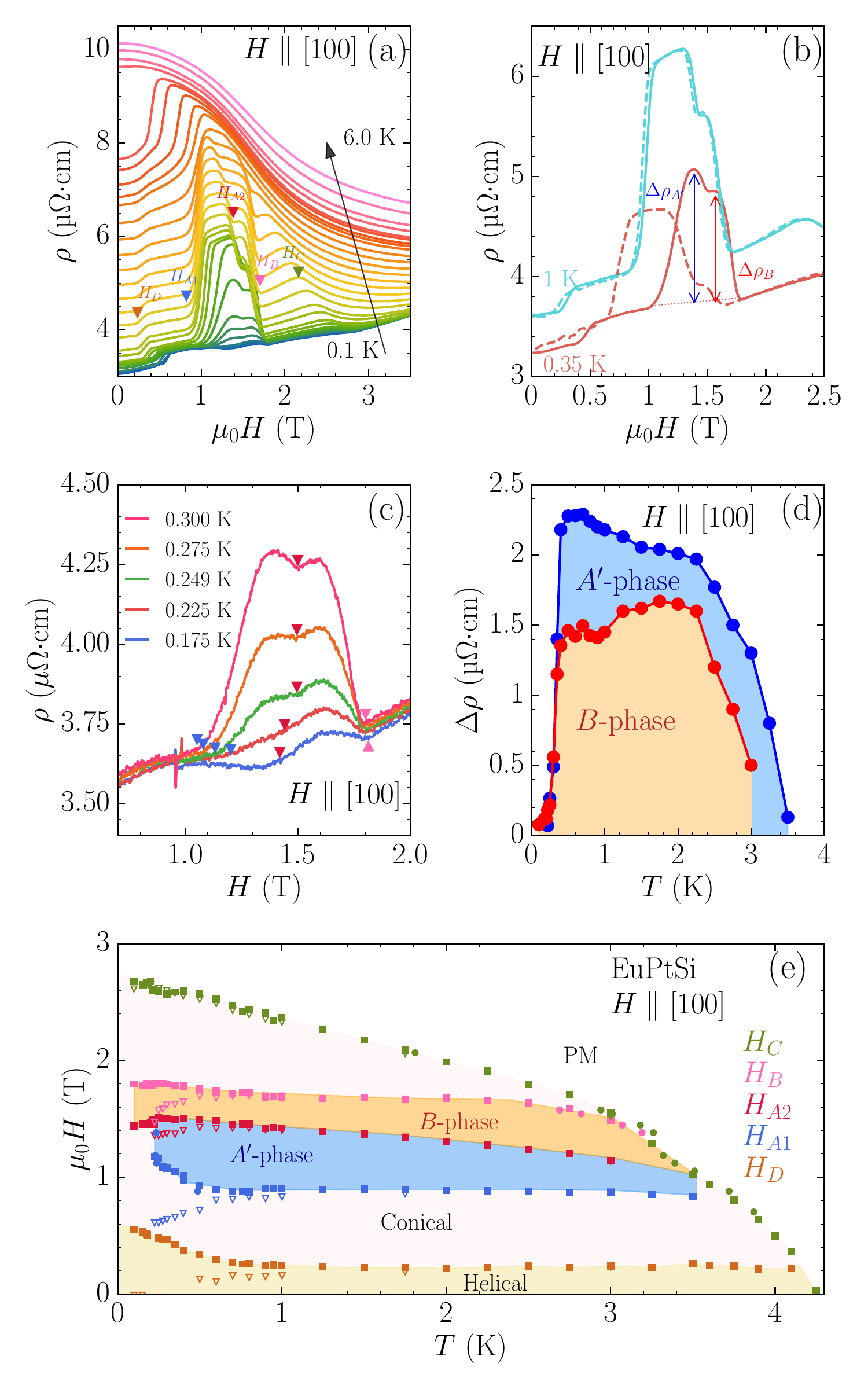}
\caption{(a) $\rho(H)$ for $H \parallel [100]$ between 0 and 3.5~T. Colored arrows denote the anomalies corresponding to transitions $H_{D}$ (orange), $H_{A1}$ (blue), $H_{A2}$ (red), $H_{B}$ (pink) and $H_{C}$ (green) in the order of increasing field. (b) $\rho(H)$  in increasing (full lines) and decreasing (dashed lines) fields for 1~K (almost no hysteresis) and 0.35~K (strong hysteresis). Also shown are the additional scattering contributions $\Delta \rho_{A'}$ and $\Delta \rho_B$ to the resistivity appearing in the A' and the B-phase, respectively. (c) $\rho(H)$ for $T \leq 0.3$~K, in the conical order between 0.7~T and 2~T. The triangles indicate the different anomalies. $H_{A1}$ disappears below 0.225~K as the A'-phase closes. $H_{A2}$ and $H_B$ are observed at all temperatures. (d) Amplitude of the contributions $\Delta \rho_{A'}$ and $\Delta \rho_B$ in the resistivity as a function of the temperature (from increasing field sweeps). (e) Magnetic phase diagram obtained for the [100] direction by resistivity measurements. Full squares (open triangles) are anomalies from increasing (decreasing) field sweeps. Full circles are anomalies from $\rho(T)$. Colored areas indicate the various magnetic structures.}
\label{Fig3}
\end{center}
\end{figure}

The field-dependence of the resistivity $\rho(H)$ for $H \parallel [100]$ for increasing magnetic field is shown in Fig.~\ref{Fig3}(a) for fields below 3.5~T and various temperatures ranging from 0.1~K to 6~K. The current has been applied along [011] $\perp H$. In the paramagnetic regime ($T > 4.05$~K), the magneto-resistance $\rho (H)$ decreases continuously. Within the magnetic ordered state below 4.05~K, the resistivity is reduced, compared to the paramagnetic state. This is clearly evidenced in $\rho(H)$ for $T=4$~K by a step-like increase at $H_C = 0.4$~T, when leaving the magnetically ordered state for the paramagnetic one (roughly recovering the same behavior as for $T > 4.05$~K). For lower temperatures the resistivity curves show several anomalies as a function of the magnetic field, which are denoted with triangles in Fig.~\ref{Fig3}(a) for the curve at $T=1.6$~K. It is noticeable that the transition from the helical to conical state at $H_D$ corresponds to an increase in $\rho (H)$, whereas $\rho(H)$ decreases at that transition for $H \parallel [111]$ (previous orientation). 
Within the conical state, $\rho(H)$ increases almost linearly and exhibits a large upturn at $H_{A1}$ (blue triangle) when entering into the A'-phase ($\rho(H)$ is enhanced by up to 50~\% at 2~K). The $H_{A2}$ transition between the A' and B-phases (red triangle) corresponds to a small decrease of the resistivity. At $H_B$ (pink triangle), the upper field limit of the B-phase, the resistivity decreases strongly, before slightly increasing again in the conical state. The transition to the paramagnetic state at $H_C$ is accompanied by a sharp (downward) kink in $\rho (H)$ and marked on the 1.6~K curve by a green triangle.

Fig.~\ref{Fig3}(b) shows the $\rho (H)$ curves for $T = 1$~K and 0.35~K for increasing and decreasing magnetic fields. While at 1~K the hysteresis between up and down sweeps is rather small, a strong hysteresis occurs at 0.35~K evidencing the first order character of the phase boundaries of the A' and the B-phase. Fig.~\ref{Fig3}(c) displays $\rho (H)$ at low temperatures. Down to 0.175~K, a tiny anomaly of the B-phase can be observed. Fig.~\ref{Fig3}(d) represents the amplitude of the additional contributions $\Delta \rho_{A'}$ and $\Delta \rho_B$ to the resistivity which occur in the A' and the B-phase, respectively, as a function of temperature. Below 3.5~K and 3.2~K the additional scattering due to the $A$' and $B$-phases strongly increases, respectively. Below 0.4~K the additional scattering contributions vanish rapidly. 

In Fig.~\ref{Fig3}(e) we finally establish the magnetic phase diagram of EuPtSi for $H \parallel [100]$ obtained from the increasing (solid symbols) and decreasing (open symbols) field sweeps. Contrarily to the [111] direction, where the A-phase disappears below 450~mK, for $H \parallel [100]$, both the A' and the B-phases exist at lower temperatures. However, we note that a strong hysteresis between the field up and down sweeps appears below 1~K (in Fig.~\ref{Fig3}(e) full squares (open triangles) stand for increasing (decreasing) field sweeps, respectively). While the A'-phase seems to be closed at low temperatures for the field up sweeps, the B-phase does clearly not close at the lowest temperature, only the additional scattering contribution $\Delta \rho_{A'}$ fades out, a tiny contribution $\Delta \rho_B$ remains. The transition field $H_{A1}$ from the conical order to the A’-phase increases as the temperature decreases, reaching about $1.3$~T at 0.225~K, where the A’-phase is no longer visible. The $H_{A2}$ transition field, on the other hand, decreases below 0.22~K, effectively enlarging the B-phase, considering that the transition field $H_B$ between the B-phase and the conical order is clearly visible at all temperatures and does not move. The presence of $H_{A2}$ and $H_B$ at all temperatures seems to indicate that the B-phase persists at lower temperatures than the A’-phase. On the contrary, for \textit{decreasing} field sweeps it is the B-phase which seems to close at low temperatures ($H_B$ drops to 1.5~T at 0.2~K), while the A'-phase remains open to low temperatures (it becomes wider as $H_{A1}$ drops from 1~T to 0.6~T). These distinct observations clearly indicate that the field history plays an important role for the low temperature behavior (see also below).


\begin{figure}[ht!]
\begin{center}
\includegraphics[width=0.48\textwidth]{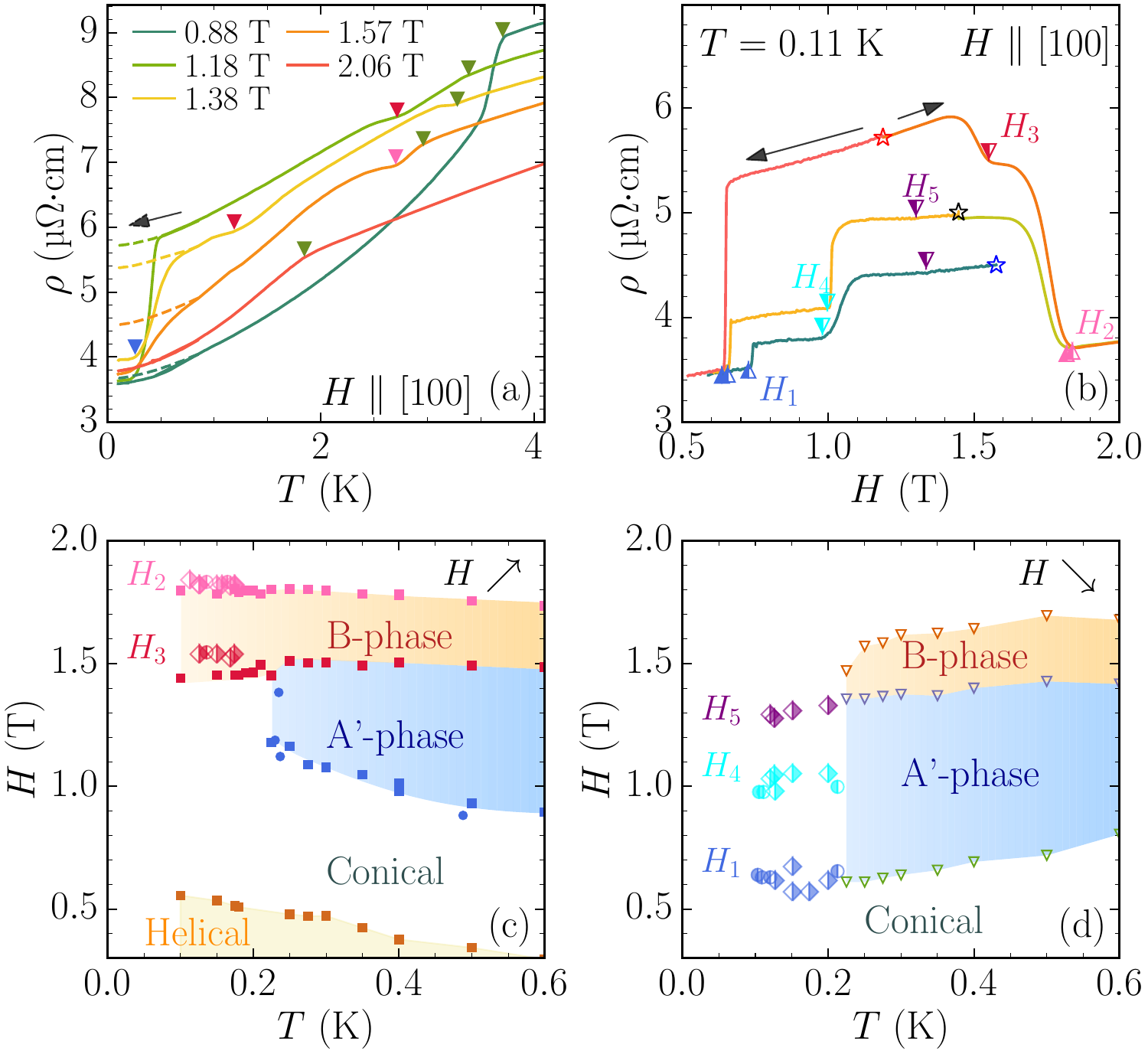}
\caption{(a) Temperature dependence of the resistivity $\rho(T)$ for $H \parallel [100]$ between 0.1~K and 4.1~K for different fields. Increasing $T$ is plotted in full lines and decreasing fields in dashed lines. 
    Colored arrows denote the anomalies corresponding to transitions $H_{A1}$ (blue), $H_{A2}$ (red), $H_{B}$ (pink) and $H_{C}$ (green). 
    (b) Field dependence of $\rho$ from the metastable A'- and B-phases induced by field cooling, for $T=0.11$~K. Stars indicate the starting point of the $H$-sweeps (respectively at 1.18, 1.35 and 1.6~T), and the direction of the sweep is given by the black arrows. Anomalies in the metastable $H$-sweeps are labeled with numbers from 1 to 5. (c) and (d) Low temperature $H-T$ phase diagram of EuPtSi with $H\parallel$~[100] for increasing and decreasing magnetic field, respectively. Full squares (open triangles) are transitions from increasing (decreasing) $H$-sweeps in the equilibrium state, and the shaded areas indicate the A'- and B-phases. 
    Full circles are transitions from (a). Anomalies in the $H$-sweeps from the metastable states are represented with half-filled markers, with the same numbering as in (b).}
\label{Fig4}
\end{center}
\end{figure}

Fig.~\ref{Fig4}(a) shows the temperature dependence of the resistivity after zero field cooling for different temperatures when sweeping across the magnetic phase diagram [see Fig.~\ref{Fig3}(e)] at different fields. For $H\approx 0.88$~T, which is just at the lower limit of the A'-phase, while heating [solid line in Fig.~\ref{Fig4}(a)] a small increase is observed at 0.8~K when entering in the A'-phase and a large step-like increase at the transition to the paramagnetic state. On cooling (dashed line) a tiny hysteresis occurs, as the system stays in the A'-phase down to the lowest temperatures. This behavior is much more pronounced at 1.18~T, where on heating a large jump of the resistivity occurs when entering the A'-phase, while on cooling no transition back to the conical phase occurs. At 1.38~T one observes the transition to the A'-phase and at higher temperatures a transition from the A' to the B-phase. At 1.57~T on heating a broader increase of the resistivity occurs when entering the B-phase, but on cooling again the transition from the B-phase back to a conical phase is supercooled, e.g the resistivity does not drop back again. Finally, at 2.06~T, no hysteresis occurs at low temperature and only a transition from the conical to the paramagnetic phase can be observed. These measurements clearly show that on cooling under magnetic field the transition from the A' and B-phases to the low temperature conical state can be supercooled and a metastable low temperature state occurs. While the transition to the A'-phase is very sharp marked by a step-like increase, the transition to the B-phase is much broader. 

Figure~\ref{Fig4}(b) shows magnetic field sweeps at constant temperature $T = 0.11$~K starting in the metastable SkL regime, established after cooling at a constant initial field. At $H=1.18$~T the sample is first in the supercooled A'-phase at $T=0.11$~K. Starting at the red star (upper curve in panel (b) of Fig.~\ref{Fig4}), by decreasing the magnetic field, a very sharp jump of the resistivity occurs at $H_1 = 0.65$~T which is of the same size than the one observed in the $T$ dependence upon warming at this field. This jump at $H_1$ corresponds to the transition from the A-phase to the low field conical phase. When increasing the field starting at $T=0.11$~K and  $H=1.18$~T, first a jump at $H_3$ indicates the transition from the A'-phase to the B-phase and another large jump at $H_2$ identifies the transition from the B-phase to the high field conical phase (see panel (c) of Fig.~\ref{Fig4}). For field sweeps starting in the supercooled B-phase (see the 2 lower curves at 1.35 and 1.6~T in panel (b) of Fig.~\ref{Fig4}, and also panel (d)), on decreasing the field a very tiny, hardly visible anomaly at $H_5$ occurs. When lowering the field further, the resistivity exhibits two more successive, yet clear-cut steps, first at $H_4$ and then at $H_1$. Remarkably, at first sight, the transition field $H_4$ does not fit into the previously found phase boundary scheme, whereas the second step at $H_1$ is consistent with the transition from the A' to the low field conical phase. When increasing the field from the metastable starting point at $T=0.11$~K and $H=1.35$~T (indicated by a star on the intermediate curve in Fig.~\ref{Fig4}(b)), a large transition occurs at $H_2$, signaling the boundary between the B and high field conical phase.

The two phase diagrams obtained from isothermal field sweeps starting in the metastable, supercooled states are shown in figs.~\ref{Fig4}(c) and (d) for increasing and decreasing field, respectively.
For field-up sweeps the transition at $H_3$ coincides with the phase line from the A' to the B-phase, while $H_2$ corresponds to the first order line from the B to the upper conical phase. For decreasing field, surprisingly three different anomalies are observed: $H_5$, $H_4$ and $H_1$. The phase diagram suggests that the line of the tiny anomaly at $H_5$ is the continuation of the transition from the B to A'-phase, but the more significant jump height of the resistivity (roughly corresponding to the one between the B and A'-phase) at $H_4$ suggests that the full transition into the A'-phase occurs only at the field $H_4$ ($<H_5$). The transition at $H_1$ can be identified as the continuation line of the transition from the A'-phase to the lower conical state.  

The important point is that even when the system is in the metastable state at low temperatures, independently whether it is a supercooled A' or B-phase, the observed sharp transitions suggest that the different spin textures of the A' and B-phase are stabilized when changing the magnetic field. This implies that the energy landscape of the different spin textures has well defined local minima which can be reached by the transition from the non-equilibrium metastable state under magnetic field. Interestingly, when starting field down sweeps in the supercooled B-phase, the transition fields seem modified and the situation more complex (observation of both $H_5$ and $H_4$) whereas the overall scheme is less affected when starting from the supercooled A'-phase.

\subsection{Angular dependence of the electronic states}

\begin{figure}[tb!]
\begin{center}
\includegraphics[width=0.48\textwidth]{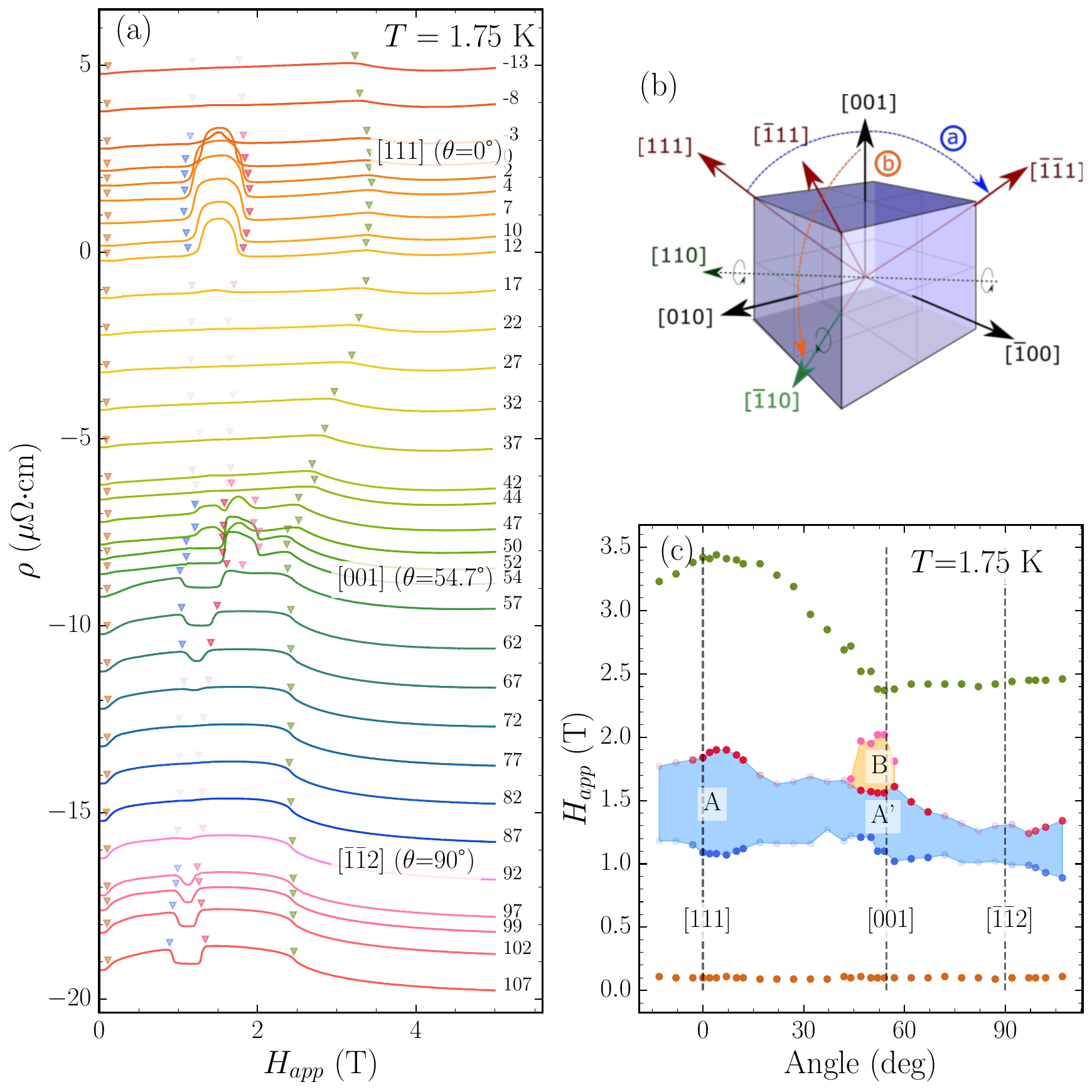}
\caption{angular dependence of the resistivity (a) $\rho(H)$ for applied fields up to 5~T at 1.75~K for various angles between $H\parallel [111] (\theta=0$°) and $H\parallel [\bar{1}\bar{1}2]$ ($\theta=90$°), sequence \textcircled{a}. Data are shifted vertically for clarity. Colored triangles indicate the respective transitions $H_D$ (brown), $H_{A1}$ (blue), $H_{A2}$ (red), $H_B$ (pink) and $H_c$ (green). (b) Simplified cubic cell indicating the principal crystallographic directions. The two rotation studies are represented by the blue arrow \textcircled{a} and the orange arrow \textcircled{b}, respectively. (c) Angle-dependent phase diagram obtained from sequence \textcircled{a}. The color strength of the transition markers correlates with the size of the anomaly in $\rho$.} 
\label{Fig8}
\end{center}
\end{figure}
Unlike MnSi and other B20 compounds, the $H$-$T$ phase diagram of EuPtSi exhibits strong anisotropy. In MnSi, the skyrmion lattice A-phase appears in a narrow region within the conical phase below $T_N$, largely independent of the magnetic field orientation. In contrast, in \mbox{EuPtSi}, the A-phase is confined to directions around $[111]$, $[\bar{1}11]$, and their equivalents. Along the $[100]$ direction, two distinct phases, A' and B, are observed, though their microscopic nature remains yet unresolved. Finally, for the $[110]$ direction, no additional phases are detected within the conical phase. Given the symmetry of the crystal structure, it is important to note that the directions $[100]$, $[010]$, and $[001]$ are equivalent, as they replicate by three-fold rotations  around the body diagonals. The lack of four-fold symmetry along the main crystallographic axes imposes a distinction between the body diagonals (the directions $[111][\bar{1}\bar{1}1][\bar{1}1\bar{1}]$ are equivalent, but not equivalent to  $[\bar{1}11][11\bar{1}][1\bar{1}1]$). 

\begin{figure}[ht!]
\begin{center}
\includegraphics[width=0.48\textwidth]{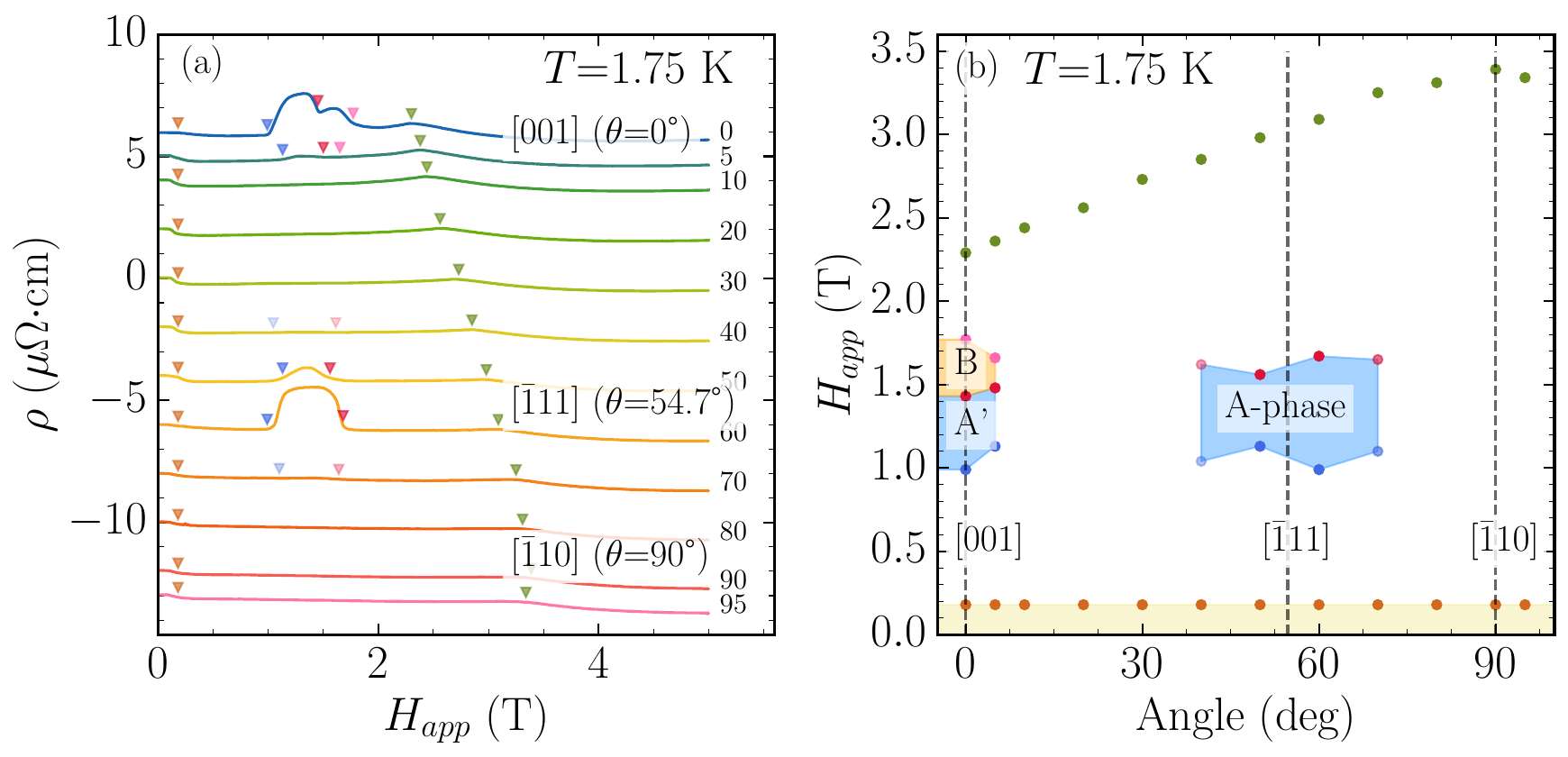}
\caption{angular dependence of the resistivity (a) $\rho(H)$ for applied fields up to 5~T at 1.75~K for various angles between $H\parallel [001] (\theta=0$°) and $H\parallel [\bar{1}10]$ ($\theta=90$°), following sequence \textcircled{b} in Fig.~\ref{Fig8}(b). Data are shifted vertically for clarity. Colored triangles indicate the respective transitions $H_D$ (brown), $H_{A1}$ (blue), $H_{A2}$ (red), $H_B$ (pink) and $H_c$ (green). (b)  Angle-dependent phase diagram obtained from sequence \textcircled{b}.}  
\label{fig9}
\end{center}
\end{figure}

The angular dependence of $\rho(H)$ at 1.75~K is shown in figs.~\ref{Fig8}(a) and \ref{fig9}(a), for applied fields up to 5~T. All data shown here stem from continuous increasing field sweeps. Between each measurement at 1.75~K, we have heated the sample to 5~K, above $T_N$, to ensure the zero-field cooling condition for the whole process. 
 For the first sequence of rotation \textcircled{a} from $H\parallel [111] (\theta=0$°) towards $H\parallel [\bar{1}\bar{1}2]$ ($\theta=90$°), the axis of rotation is the [$\bar{1}10$] direction. The rotation sequence is schematically represented in Fig.~\ref{Fig8}(b) by a dashed blue arrow. The rotation axis is shown in green and the main crystallographic axes are highlighted in red color. 

For all angles, at zero field, the system is in the helical state until the transition $H_D$ occurring around 0.1~T. Between $H_D$ and $H_C$, the system is supposedly in the conical order. Above $H_C$, the moments are oriented along the applied magnetic field direction and the system is in the field-polarized paramagnetic state.

The A-phase visibly exists from $\approx$ -3° to $\approx$ 15° around the $H\parallel$~[111] direction.
In this angular range, the resistivity is strongly enhanced by the additional scattering due to the skyrmion lattice. Since the demagnetizing field correction depends on the sample geometry with respect to field, we show the resistivity as a function of the applied field. As EuPtSi has a cubic structure, the two equivalent diagonals $[111]$ and $[\bar{1}\bar{1}1]$ are separated by 109.4°, which explains the presence of the A-phase at 95° and beyond. For those large angles, the field is applied mostly along the current direction $[\bar{1}\bar{1}2]$ and the skyrmion lattice phase corresponds to a decrease in resistivity, which is consistent with the measurements from Takeuchi \textit{et al.}\cite{takeuchi_angle_2020} This can be explained by the relative angle between the $q$-vectors of the ordered phases and the current direction. Since the population of domains for different $q$-vectors changes along the field direction above $H_D$, the variation of angle between the field and the current can even change the sign of the specific resistivity contribution ($q$-dependent scattering) in the ordered phases.

For angles between 45$\degree$ and 58$\degree$, $\rho(H)$ reveals the emergence of two distinct phases. In the range $45\degree < \theta < 50\degree$, both phases appear as positive anomalies in the resistivity (additional scattering). Between $50\degree < \theta < 58\degree$, the contribution of the lower-field phase becomes negative, while the higher-field phase vanishes for $\theta > 58\degree$. These two features correspond to the A'- and B-phases near the $H \parallel [100]$ direction. As indicated by the colored arrows and markers in Fig. \ref{Fig8}(a) and (c), the transitions associated with the A-phase are observed continuously across the full range of studied angles. Even if the transition becomes very subtle (on the order of $\approx 0.01~\mu\Omega\text{cm}$ at 32°), it remains detectable. For this specific rotation sequence (\textcircled{a}), the A- and the A'-phase appear to be connected.

A second angular dependence, sequence \textcircled{b} (schematically represented in Fig.~\ref{Fig8}(b)), has been performed based on the equivalent rotational axis $[110]$ (as the current direction) from $H\parallel [001]$ to $[\bar{1}10]$, yielding the results shown in Fig.~\ref{fig9}. The A-phase is only observed for $40\degree < \theta < 70\degree$ and is not connected to the A'-phase, the latter being present for $\theta < 10\degree$. The B-phase is observed only within a 5° angular range close to $H\parallel [001]$.
Due to the particular symmetries and chirality of the crystal structure of EuPtSi, the A- and A'-phases appear to be continuously connected for certain (rotational) axes, and completely separated for others.

\subsection{Topological Hall effect}

\begin{figure}[t!]
\begin{center}
\includegraphics[width=0.48\textwidth]{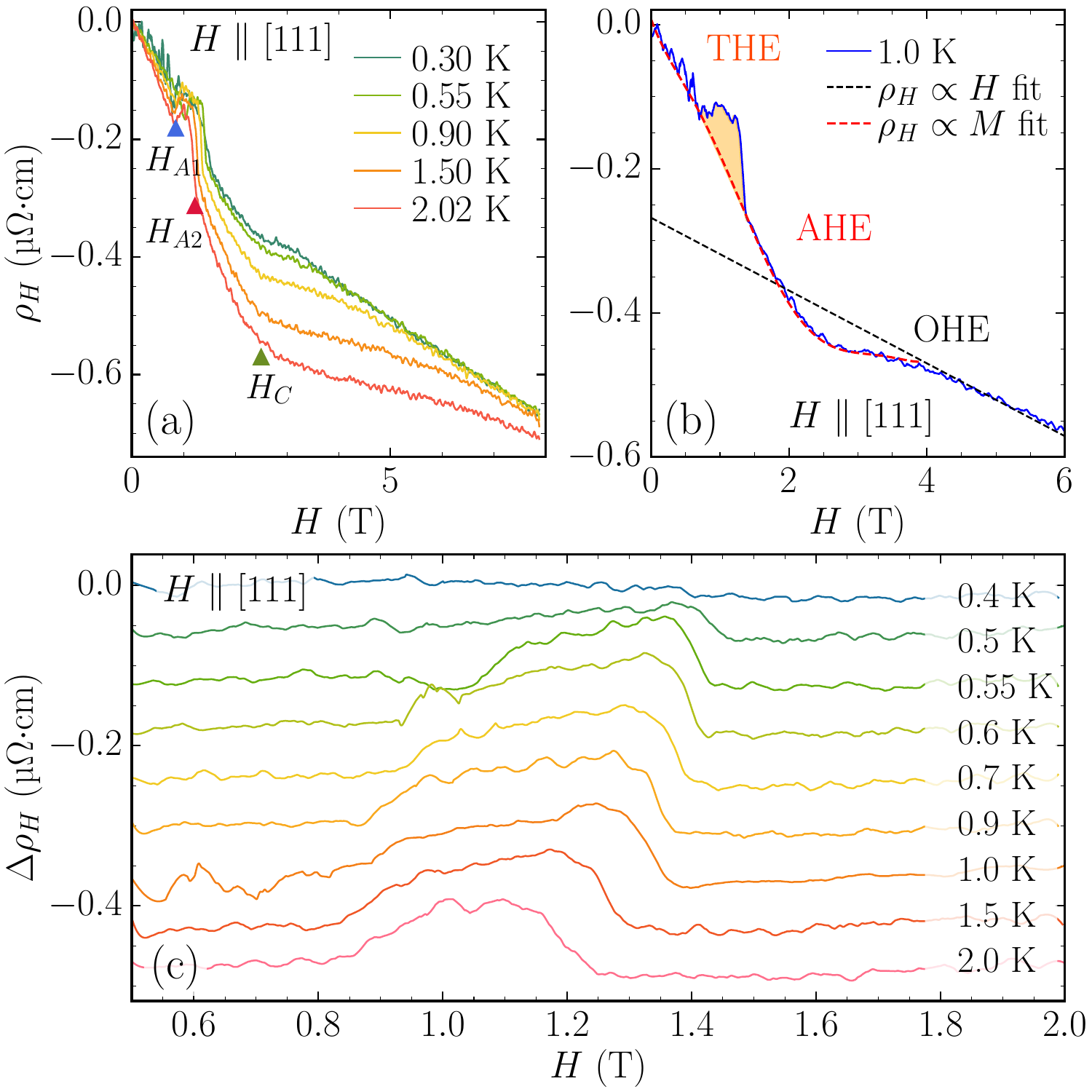}
\caption{(a) Hall resistivity $\rho_H$ as a function of magnetic field applied along [111] for various temperatures, triangles indicate  the position of the transitions to the A-phase, and the critical field $H_C$ for the curve at $T=2$~K. (b) For the $\rho_H$ data at $T = 1$~K the three different contributions to the Hall effect are indicated, the ordinary Hall effect OHE (dashed black line), the anomalous Hall effect AHE (dashed red line) and the topological Hall effect THE (orange area), see text. (c)~Spotlight on the THE contribution as a function of magnetic field for different temperatures, curves are shifted vertically for clarity (a linear contribution obtained from a fit at low field, mainly reflecting the AHE (see for example the dashed red line in (b) at 1~K), has been subtracted from the original $\rho_H(H)$ data).}
\label{Fig5}
\end{center}
\end{figure}

A hallmark of the topological nature of the skyrmion phases is the observation of the topological Hall effect (THE).\cite{neubauer_hall_2009} The field dependence of the Hall resistivity $\rho_H$ is presented in Fig.~\ref{Fig5}(a) for various temperatures and field applied along [111]. The anomalies corresponding to $H_{A1}$, $H_{A2}$ and $H_{C}$ are highlighted for $T=2$~K. The skyrmion A-phase is clearly evidenced by an additional contribution to the Hall resistivity forming a peak structure between $H_{A1}$ and $H_{A2}$ (see figs.~\ref{Fig5}(b) and (c)). In order to pinpoint the contribution of the skyrmion lattice, we assume that the Hall resistivity is given by the sum of three different contributions\cite{CULCER2024587}, see Fig.~\ref{Fig5}(b): $\rho_H=\rho_{\text{OHE}}+\rho_{\text{AHE}}+\rho_{\text{THE}}$, where $\rho_{\text{OHE}}=R_{\text{OHE}}B$ is the ordinary contribution proportional to the magnetic field (and inversely proportional to the carrier density, $R_{\text{OHE}}=\frac{1}{n_e}$), $\rho_{\text{AHE}}$ is the anomalous contribution primarily driven by spin-orbit interactions, and $\rho_{\text{THE}}$ is the topological contribution in magnetic systems where a non-zero scalar spin chirality is present. The anomalous part which is proportional to the magnetization $M$ of the system ($\rho_{\text{AHE}}=R_SM$) can occur through three main mechanisms (on which the detailed expression for $R_S$ depends): intrinsic mechanism (Berry phase contribution) related to the Berry curvature of the electronic band structure ($\rho_{\text{AHE}} \propto M\rho^2$), skew scattering related to the asymmetry in electron scattering off impurities ($\rho_{\text{AHE}} \propto M\rho$), or side jump mechanism where electrons experience an additional lateral displacement when scattering off impurities (the latter mechanism only plays a supposedly minor role here) \cite{PhysRevResearch.2.033179}.

The question whether the anomalous part of the Hall effect is intrinsic or extrinsic is still open in EuPtSi. First because of the weak field dependence of the longitudinal resistivity $\rho$ at low field and the temperatures concerned by the magnetic phases, i.e. in a region where the magnetization increases strongly with field, so that the experimental distinction between the two regimes $\rho_{\text{AHE}}\propto M\rho^2$ or $\rho_{\text{AHE}}\propto M\rho$ is not straightforward. Second because the skew scattering process gets scattering-free below a certain coherence temperature (like in a Kondo lattice) \cite{PhysRevResearch.2.033179}, modifying its contribution from $\rho_{\text{AHE}}\propto M\rho$ (above the coherence temperature) to $\rho_{\text{AHE}}\propto M\rho^2$, similar to the one expected for the intrinsic Berry phase contribution, hindering again a clearcut experimental differentiation. 

The OHE is extracted from the high field data $H > H_C$, where the magnetization is saturated, e.g. above 4~T using a linear fit [see black dashed line in Fig.~\ref{Fig5}(b)]. We obtain an ordinary Hall constant $R_{\text{OHE}}= \rho_{\text{OHE}}/B= 0.051~\mu\Omega$cm/T. The corresponding carrier concentration, assuming only one band at the Fermi level, yields $n_e=1.23 \times 10^{22}$~cm$^{-3}$, which is within the expected range for a metal. At lower field we can assume that the Hall effect is mainly governed by the AHE contribution (red dashed line in Fig.~\ref{Fig5}(b)) because of the large increase of the magnetization in this field range. We can finally extract the contribution to the Hall effect due to the presence of the SKL phase by fitting $\rho_H(H)$ linearly between 0.5~T and 2~T in the conical phase, thus removing the OHE and the dominant AHE contribution, see Fig.~\ref{Fig5}(c). The only remaining (field dependent) contribution is the THE due to the SKL in the A-phase, which reaches a maximum of $\Delta \rho_{H\_A} = 0.1~\mu\Omega$cm at 0.6~K. The disappearance of the anomaly below 0.5~K is in good agreement with the phase diagram previously measured.
 
 The observed THE arises from the deflection of conduction electrons by the emergent magnetic field generated by the SKL in the A-phase. This is distinct from the Skyrmion Hall effect, which results from the motion of skyrmions themselves under applied electric and magnetic fields. The THE signal can be expressed as $\Delta \rho_{\text{THE}} \approx P\cdot R_{\text{OHE}}\cdot B_{\text{em}}$ with $P$ the local spin polarization of the conduction electrons and $B_{\text{em}}$ the emergent magnetic induction.\cite{neubauer_hall_2009} $P$ is maximum ($P=1$) for a fully polarized system and minimum ($P=0$) for vanishing polarization. In between $P$ is given by the ratio of the ordered magnetic moment in the A-phase $\mu_{A}=0.7~\mu_B$ with respect to the saturated magnetic moment $\mu_{\text{sat}}=7~\mu_B$, $P=\frac{\mu_{A}}{\mu_{\text{sat}}}=0.1$. 
The emergent field $H_{\text{em}}$ originates form the Berry curvature associated with the nontrivial spin configuration of skyrmions. Electrons traveling through such a texture pick up a Berry phase, which behaves as if the electrons were in a magnetic field. The associated magnetic flux is given by the surface integral $\Phi_{\text{em}}= \iint B_{\text{em}}dS=\Phi_0 N_{\text{sky}}$ where $N_{\text{sky}}$ is the so-called skyrmion number (topological charge), typically $\pm 1$ per skyrmion ($-1$ for an antiskyrmion) and $\Phi_0=\frac{h}{e}$ the flux quantum. $B_{\text{em}}=-\frac{\Phi_0}{A_{\text{sky}}}$ where $A_{\text{sky}}$ is the area occupied by a single skyrmion, with the negative sign indicating that $B_{\text{em}}$ opposes the applied magnetic field. In EuPtSi the skyrmion size is small, $\lambda_{\text{sky}}= 18$~\AA\ (from the periodicity of the helical state near $T_N$), and taking into account the hexagonal structure of the skyrmion lattice, we have $A_{\text{sky}}=\frac{2 \lambda_{\text{sky}}^2}{\sqrt3}$, giving $B_{\text{em}}=-1105$~T. Because the flux is packed into a very small real-space area (of the order of the skyrmion size), the effective emergent magnetic induction is huge. For comparison, in MnSi due to the larger size of the skyrmions $B_{\text{em}}=-13.5$~T.\cite{neubauer_hall_2009} Even though $B_{\text{em}}$ is extremely large (compared to the upper field limit of the SKL phase) in both systems, the resulting THE is relatively small compared to the conventional Hall effect observed under much weaker applied magnetic fields. This can be attributed to two main reasons: First, the average emergent field across the entire sample is much smaller than the local peak value, because most of the material contains either no skyrmions or a diluted spin texture. Hall measurements reflect this sample-averaged response, not the localized maximum field within each skyrmion. Second, only a small fraction of conduction electrons contribute to the THE. Specifically, it involves electrons that couple to the spin texture and are sensitive to the real-space Berry curvature. Many electrons — such as those with opposite spin orientation or those undergoing incoherent scattering — do not effectively contribute. The Berry phase effect is highly selective, meaning not all electrons participate equally in generating the Hall voltage.

\begin{figure}[ht!]
\begin{center}
\includegraphics[width=0.45\textwidth]{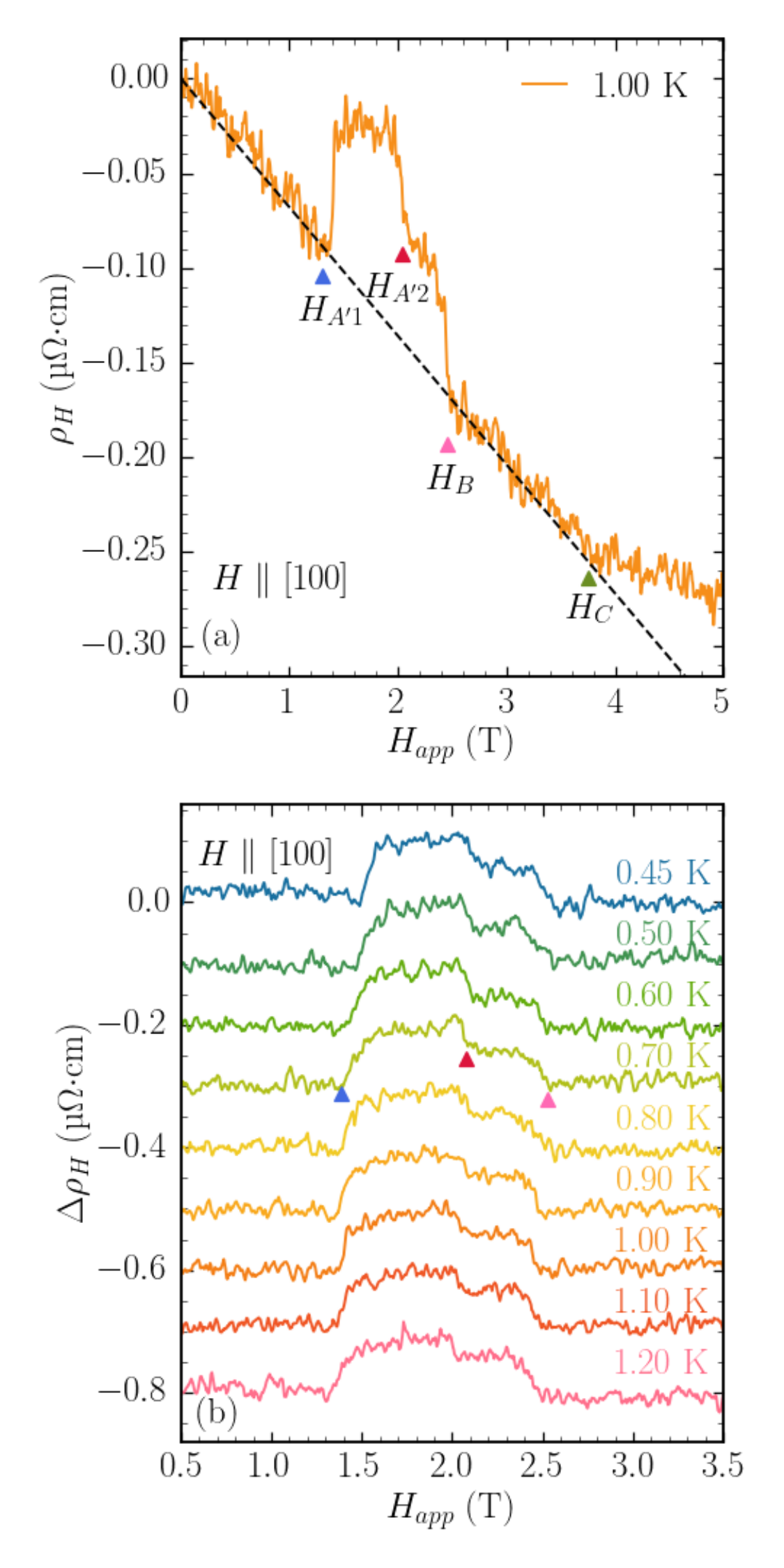}
\caption{(a) Hall resistivity $\rho_H$ as a function of magnetic field applied along [100] for various temperatures, triangles indicate  the position of the transitions to the A'- and B-phase, and the critical field $H_C$. (b) Contribution of the topological Hall resistance as a function of magnetic field for different temperatures, curves are shifted vertically for clarity (same extraction method as for field along [111]).}
\label{Fig6}
\end{center}
\end{figure}
We also have performed Hall resistivity measurements for $H \parallel [100]$ where two phases A' and B are present, see Fig.~\ref{Fig6}(a). The anomalies associated with the transitions $H_{A1}$, $H_{A2}$, $H_B$, and $H_C$ are marked by colored triangles. The A'-phase is characterized by a pronounced positive contribution, similar to that observed in the A-phase for field along [111]. Likewise, the transition into the B-phase from the conical phase is marked by a smaller positive contribution. Following the same analysis as used for $H \parallel [111]$, the isolated THE is presented in Fig.~\ref{Fig6}(b). The A'- and B-phase yield a maximum contribution of $\Delta \rho_{H\_A'} = 0.1~\mu\Omega$cm at 0.7~K and $\Delta \rho_{H\_B} = 0.07~\mu\Omega$cm at 0.8~K, respectively. In contrast to the A-phase, neutron scattering experiments have not yet resolved the microscopic structures of the A'- and B-phase, whether they are skyrmionic, multi-${\mathbf q}$ or single-${\mathbf q}$ in nature. Without a clear understanding of the underlying magnetic topological order, it is not possible to extract $B_{\text{em}}$ from the data. Nevertheless, the observation of a positive contribution to the Hall effect strongly points towards a topological origin for both the A'- and the B-phase, as suggested by other previously shown similarities (metastability, etc.) with the A-phase.

\section{Conclusion}

In this work, we have systematically investigated the magnetic phase diagram of the chiral antiferromagnet EuPtSi through resistivity and Hall effect measurements under various magnetic field orientations. We confirmed the existence of a skyrmion lattice (SkL) A-phase for fields applied along the [111] direction and demonstrated its remarkable metastability down to the lowest temperatures via field cooling. Notably, this metastability is independent on the cooling rate and magnetic history, and it remains stable over extended periods, highlighting the robustness of the topologically protected state.  However, understanding and controlling the energy barriers that govern the skyrmion lattice creation and annihilation remains a key challenge for a further exploration of the metastable behavior.

Moreover, we identified and characterized two additional topological magnetic phases  — the A'- and B-phase — for fields along the [100] direction. These phases also exhibit metastability and hysteretic behavior, suggesting similar underlying topological protection. While their microscopic nature remains unresolved, the observation of topological Hall signals in these phases supports a topologically nontrivial spin texture.
The angular dependence of the resistivity further reveals strong anisotropy in the magnetic phase diagram, highlighting the role of the crystallographic orientation in stabilizing distinct topological states. Our findings position \mbox{EuPtSi} as a unique platform where quantum and thermal effects interplay to stabilize multiple metastable topological magnetic phases, opening new avenues for exploring quantum skyrmionics in centrosymmetric and antiferromagnetic systems.




\bibliography{references}


\end{document}